# Relationship Between Phonons and Thermal Expansion in $Zn(CN)_2$ and $Ni(CN)_2$ from Inelastic Neutron Scattering and Ab-Initio Calculations

R Mittal[1], M. Zbiri[2], H. Schober[2,3], E Marelli[4], S. J. Hibble[4], A. M. Chippindale[4] and S. L Chaplot[1]

[1]Solid State Physics Division, Bhabha Atomic Research Centre, Trombay, Mumbai 400 085, India
[2]Institut Laue-Langevin, BP 156, 38042 Grenoble Cedex 9, France
[3]Université Joseph Fourier, UFR de Physique, 38041, Grenoble Cedex 9, France
[4]Department of Chemistry, University of Reading, Whiteknights, Reading, Berks RG6 6AD, UK

$Zn(CN)_2$ and $Ni(CN)_2$ are known for exhibiting anomalous thermal expansion over a wide temperature range. The volume thermal expansion coefficient for the cubic, three dimensionally connected material, $Zn(CN)_2$, is negative ($\alpha_V = -51 \times 10^{-6}$ K$^{-1}$) while for $Ni(CN)_2$, a tetragonal material, the thermal expansion coefficient is negative in the two dimensionally connected sheets ($\alpha_a = -7 \times 10^{-6}$ K$^{-1}$), but the overall thermal expansion coefficient is positive ($\alpha_V = 48 \times 10^{-6}$ K$^{-1}$). We have measured the temperature dependence of phonon spectra in these compounds and analyzed them using *ab-initio* calculations. The spectra of the two compounds show large differences that cannot be explained by simple mass renormalization of the modes involving Zn (65.38 amu) and Ni (58.69 amu) atoms. This reflects the fact that the structure and bonding are quite different in the two compounds. The calculated pressure dependence of the phonon modes and of the thermal expansion coefficient, $\alpha_V$, are used to understand the anomalous behavior in these compounds. Our *ab-initio* calculations indicate that it is the low-energy rotational modes in $Zn(CN)_2$, which are shifted to higher energies in $Ni(CN)_2$, that are responsible for the large negative thermal expansion. The measured temperature dependence of the phonon spectra has been used to estimate the total anharmonicity of both compounds. For $Zn(CN)_2$, the temperature- dependent measurements (total anharmonicity), along with our previously reported pressure dependence of the phonon spectra (quasiharmonic), is used to separate the explicit temperature effect at constant volume (intrinsic anharmonicity).

**I. INTRODUCTION**

The discovery of anomalous thermal expansion in framework solids is of fundamental scientific interest and may find application in fabricating technological materials, in particular for the optical and electronics industries. Besides oxide-based materials [1-2], anomalous thermal expansion behavior has been observed in molecular framework materials containing linear diatomic bridges such as the cyanide anions [3-6]. Recent X-ray diffraction measurements have shown [6] that the hexagonal materials, $Ag_3Co(CN)_6$ and $Ag_3Fe(CN)_6$, exhibit exceptionally large ("colossal") positive thermal expansion (PTE) along the *a* direction ($\alpha_a = +140 \times 10^{-6}$ K$^{-1}$) and negative thermal expansion (NTE) along the *c* direction ($\alpha_c = -125 \times 10^{-6}$ K$^{-1}$). These thermal expansion coefficients are an order of magnitude larger than those observed in any other material. Even simple cyanides such as $Zn(CN)_2$ are reported [5] to have an isotropic NTE coefficient ($\alpha_V = -51 \times 10^{-6}$ K$^{-1}$), which is twice as large as that of $ZrW_2O_8$ [1]. However, when Zn is substituted by Ni, a layered compound, $Ni(CN)_2$, is produced [3] which has NTE in two dimensions ($\alpha_a = -7 \times 10^{-6}$ K$^{-1}$) combined with a very large positive (PTE) coefficient ($\alpha_c = 61.8 \times 10^{-6}$ K$^{-1}$) in the third dimension perpendicular to the layers, to yield a large overall volume thermal expansion ($\alpha_V = 48.5 \times 10^{-6}$ K$^{-1}$).

It has been proposed from pair distribution function (PDF) analysis of the structural data collected using high energy X-rays that NTE in $Zn(CN)_2$ is induced by an average increase of the transverse thermal amplitude of the motion of bridging C/N atoms, away from the body diagonal [7]. Further, investigation using *ab-initio* calculations [8] of the geometry and electronic structure of $Zn(CN)_2$ shows that the naturally stiff C≡N bond is paired with weak Zn–C/N bonds. This type of bonding allows large transverse thermally excited motions of the bridging C/N atoms to occur in (M-CN-M) bridges within metal-cyanide frameworks. Structural studies [9] of $Zn(CN)_2$ show that two different models having cubic symmetry with space group *Pn3m* (disordered model) and *P43m* (ordered model), give equally good account of the diffraction data. The ordered structure (Fig. 1) consists of a $ZnC_4$ tetrahedron (at the centre of the cell) linked to four neighboring $ZnN_4$ tetrahedra (at the corners of the cell) with CN groups along four of the body-diagonals. In the X-ray diffraction modelling of the disordered structure, atomic sites are given 50:50 C:N occupancy. Such models cannot be used for *ab-initio* calculations where sites can contain only one type of atom, thus restricting the space group to *P43m*.

Ni(CN)$_2$ is fundamentally different from Zn(CN)$_2$, in that it forms a layered structure with average tetragonal symmetry. Nickel cyanide has a long-range ordered structure in two dimensions (*a-b* plane) (Fig. 2) but a high degree of stacking disorder in the third dimension. The relationship between neighbouring layers is defined, but there is a random element to the relationship between next nearest neighbours. A crystallographic model in *P4$_2$/mmc* [3] reproduces the structure well and the disorder in the stacking is dealt with more comprehensively in the paper by Goodwin *et al.* [4]. The covalent bonding within the layers is much stronger than van der Waals' bonding between the layers, and it is this component that is studied in the calculations presented here, when effectively isolated Ni(CN)$_2$ layers are considered. Nickel cyanide shows anisotropic thermal behavior. Although the dimensions of its square grid like layers (*a-b* plane) decrease with increasing temperature, this decrease is accompanied by an length increase along the *c* direction giving an overall positive thermal expansion.

As already mentioned above, motions at right angles to the atomic and/or molecular bonds are identified as the principal cause of anomalous thermal expansion in framework compounds. Such motions are necessarily connected with transverse vibrations. For the amplitudes of motion to be large, the corresponding vibrations should be low in energy. That low-energy phonon modes play an important role in anomalous thermal expansion has been demonstrated by previous work [10-15] on ZrW$_2$O$_8$ and HfW$_2$O$_8$. In the case of Zn*(CN)*$_2$, time-of-flight inelastic neutron scattering measurements from powdered samples [16] indicate the existence of dispersionless modes at about 2 meV (~16 cm$^{-1}$). To produce thermal expansion, vibrations not only have to be of large amplitude, but also have to be anharmonic in nature. To this end, we have recently investigated the anharmonicity of phonons in Zn(CN)$_2$ [17] by employing a high-pressure inelastic neutron scattering technique. In this article we extend these investigations by including the temperature dependence of the Zn(CN)$_2$ spectra, together with a comparison with Ni(CN)$_2$. The analysis of the experiments is performed with the help of state-of-the-art *ab-initio* lattice dynamical calculations. In this way we obtain a clear and detailed insight into the phonon mechanisms responsible for thermal expansion in Zn(CN)$_2$ and Ni(CN)$_2$.

## II. EXPERIMENTAL

Zn(CN)$_2$ (~98.0% pure) polycrystalline sample was obtained from Aldrich, USA. Ni(CN)$_2$.1.5H$_2$O, purchased from Alfa Aesar, was dried under vacuum at room temperature for 12 hours reground and then dried under vacuum at 200$^o$ C for three weeks. Powder X-ray diffraction showed that the hydrated nickel cyanide had been completely converted to anhydrous nickel cyanide and that Ni(CN)$_2$ was the only crystalline phase present. The IR spectrum of Ni(CN)$_2$, collected using a Perkin

Elmer 100 FT-IR spectrometer with Universal ATR Sampling accessory, confirmed that the material was fully dehydrated. The Raman spectrum was collected on a Renishaw InVia Raman microscope.

The inelastic neutron scattering experiments were performed using the IN6 time-of-flight spectrometer at the Institut Laue-Langevin (ILL), in Grenoble, France. The temperature-dependent measurements were performed on about 10 grams of polycrystalline samples of $Zn(CN)_2$ and $Ni(CN)_2$. The samples were placed in a cryostat inside sealed thin-slab aluminum containers mounted at 45° with respect to the incident neutron beam. The high-resolution data for $Zn(CN)_2$ and $Ni(CN)_2$ were measured at several temperatures from 300 K to 160 K. The measurements were performed in the neutron-energy gain inelastic focusing mode with an incident neutron wavelength of 5.12 Å (3.12 meV) and 4.14 Å (4.77 meV) for $Zn(CN)_2$ and $Ni(CN)_2$, respectively,. The energy resolution of the spectrometer is 0.20 meV for λ=5.12 Å and 0.30 meV for λ=4.14 Å, in the inelastic focusing mode. The spectrometer is equipped with a large detector bank covering a wide range of scattering angle (10° to 115°).

In the incoherent one-phonon approximation, the phonon density of states [18] is related to the measured scattering function $S(Q,E)$, as observed in the neutron experiments by:

$$g^{(n)}(E) = A < \frac{e^{2W_k(Q)}}{Q^2} \frac{E}{n(E,T) + \frac{1}{2} \pm \frac{1}{2}} S(Q,E) > \qquad (1)$$

$$g^n(E) = B \sum_k \{\frac{4\pi b_k^2}{m_k}\} g_k(E) \qquad (2)$$

where the + or – signs correspond to energy loss or gain of the neutrons, respectively, and $n(E,T) = [\exp(E/k_BT) - 1]^{-1}$. $A$ and $B$ are normalization constants and $b_k$, $m_k$, and $g_k(E)$ are, respectively, the neutron scattering length, mass, and partial density of states of the $k^{th}$ atom in the unit cell. The quantity within < ---- > represents an appropriate average over all $Q$ values at a given energy. $2W(Q)$ is the Debye-Waller factor. The weighting factors $\frac{4\pi b_k^2}{m_k}$ for each atom type in the units of barns/amu are: Zn: 0.063; Ni: 0.315; C: 0.462 and N: 0.822 calculated from the neutron scattering lengths found in Ref. [19].

## III. LATTICE DYNAMICAL CALCULATIONS

*Ab-initio* calculations were performed using the projector-augmented wave (PAW) formalism [20] of the Kohn-Sham DFT [21, 22] at the generalized gradient approximation level (GGA), implemented in the Vienna *ab initio* simulation package (VASP) [23, 24]. The GGA was formulated by the Perdew-Burke-Ernzerhof (PBE) [25, 26] density functional. The Gaussian broadening technique was adopted and all results are well converged with respect to *k*-mesh and energy cutoff for the plane wave expansion. The partially relaxed (only coordinates are optimized) ordered structures of $Zn(CN)_2$ and $Ni(CN)_2$ were used for the lattice dynamical calculations and related properties. For $Zn(CN)_2$, the available structure having the cubic space group (*P43m* (215), $\left[T_d^1\right]$) is considered [9]. For $Ni(CN)_2$, a periodic model system is used (Table I) to generate the layers within the tetragonal space group (*P4/mmm* (75) $\left[C_4^1\right]$) [4]. This model for $Ni(CN)_2$ is an approximation of the real situation as the interlayer spacing used is double that found in the actual material. Such a model results in no interaction between the layers. However, the model can be used to reproduce most features of the Raman spectrum, the phonon density of states (DOS) and, in addition, can be used to investigate the in-plane negative thermal expansion.

In the lattice dynamics calculations, in order to determine all interatomic force constants, the supercell approach has been adopted [27]. For both $Ni(CN)_2$ and $Zn(CN)_2$, (2*a*, 2*b*, 2*c*) supercells containing 16 formula units (80 atoms) were constructed. Total energies and interatomic forces were calculated for the 20 structures generated for $Ni(CN)_2$ and for the 8 structures, generated for $Zn(CN)_2$, by displacement of the four symmetry inequivalent atoms present in both systems along the three Cartesian directions ($\pm x$, $\pm y$ and $\pm z$). Phonon density of states (PDS), phonon dispersion relations (PDR) and Raman active modes/frequencies were extracted in subsequent calculations using the Phonon software [28].

## IV. RESULTS AND DISCUSSION

### A. Phonon density of states and dispersion relation

The measured temperature dependence of the phonon spectra for $Zn(CN)_2$ and $Ni(CN)_2$ are shown in Figs. 3 and 4, respectively. Differential scanning calorimetric measurements [29] for $Zn(CN)_2$ show that there is an order-disorder transition at about 250 K. We have measured phonon spectra (Fig.

3) at 180, 240, 270 and 320 K. Our measurements show that there is no significant change in the phonon spectra. for either compound as a function of temperature. All the observed features for Zn(CN)$_2$ are nicely reproduced computationally, especially the low-energy peak at ~ 2 meV. This means that the order-disorder transition in Zn(CN)$_2$ is only a weak perturbation to the vibrational system.

The experimental phonon spectrum of Ni(CN)$_2$ (Fig. 4) shows several well-pronounced vibration bands. The calculated positions of these bands match very well with the experimental data while there are slight differences in the intensities. This is probably due to the fact the interlayer interactions in Ni(CN)$_2$ have not been included. As we do neglect the interlayer coupling, the modes along the stacking axis have very low energies. All these modes are included in the calculated density of states that we have shown in Fig. 4. This explains the extra weight in the calculated density of states at low energies.

The comparison of the phonon spectra of Zn(CN)$_2$ and Ni(CN)$_2$ shows (Fig. 5) that there are pronounced differences. The cut-off energy for the external modes in Zn(CN)$_2$ and Ni(CN)$_2$ is at about 65 meV and 90 meV, respectively The calculated partial density of states show that the contributions from Zn and Ni in Zn(CN)$_2$ and Ni(CN)$_2$ (Fig. 6) extend up to 60 and 75 meV, respectively. These can obviously not be explained by a simple mass renormalization of the modes involving Zn (65.38 amu) and Ni (58.69 amu) atoms. They thus imply that the strength and may be the character of bonding is different in both systems. Our inelastic scattering data show that the first low-energy band in the Zn compound is at 2 meV, while in the Ni compound this band is at 10 meV.

The structures of Zn(CN)$_2$ and Ni(CN)$_2$ yield 30 phonon modes for each wave vector. The comparison between the experimental and calculated zone centre modes for Zn(CN)$_2$ [30] and Ni(CN)$_2$ is given in Tables II and III, respectively. The agreement is very close in each case. The Raman spectrum of Ni(CN)$_2$ (Fig. 7) has not previously been reported. Fig. 8 shows the $v_{C\equiv N}$ region for both the Raman and infrared spectra. The presence of two $v_{C\equiv N}$ absorptions in the Raman and one in the infrared (Fig. 8) is consistent with $D_{4h}$ symmetry.

To our knowledge, no measurements of the phonon spectrum have been reported for Ni(CN)$_2$. However, five dispersion less phonon modes below 1.0 THz (4.136 meV) arising from motions of a single Ni(CN)$_2$ layer were found from Reverse Monte Carlo fitting of total neutron diffraction data [4].

Our calculations for Ni(CN)$_2$ do not confirm this prediction, but identify the lowest optic mode at 99 cm$^{-1}$ (~3 THz, ~12.3 meV) in agreement with the experimental data (Fig. 4).

The calculated phonon dispersion curves for Zn(CN)$_2$ are shown in Fig. 9. We notice a remarkable flat phonon dispersion sheet of the two lowest energy acoustic modes at about 2 meV. These flat modes give rise to the observed first peak in the density of states at about 2 meV. Further flat phonon dispersion sheets are found at relatively high energies of about 25, 30, 40 and 60 meV (Fig. 9) providing the other well isolated bands in the phonon density of states (Fig. 3). The calculated dispersion relation for Ni(CN)$_2$ (Fig. 9) is quite different as compared to Zn(CN)$_2$. As explained in Section III, the calculations for Ni(CN)$_2$ are carried out with the layers separated, which is different from the real situation. Further, we find that at zone boundary the acoustic modes extend up to 10 meV. The flattening of acoustic modes around 10 meV gives rise to the first peak in the density of states (Figs. 4 and 5) of Ni(CN)$_2$. The large difference in the energies of acoustic modes between the compounds indicates that bonding is quite different in both the compounds. Further, the flat phonon dispersion sheets at about 18, 30, 45, 60 and 75 meV give rise to the isolated peaks in the density of states of Ni(CN)$_2$. The Bose factor corrected S(Q,E) plots for Zn(CN)$_2$ and Ni(CN)$_2$ at 180 K and 160 K respectively are shown in Fig. 10. The figure clearly shows the presence of flat acoustic mode at 2 meV in the S(Q,E) plot of Zn(CN)$_2$, while for Ni(CN)$_2$ the acoustic modes extends up to about 10 meV.

In case of Ni(CN)$_2$, the acoustic dispersion within the sheets for the transverse branches possesses in the calculation an anomalous dispersion. The curves turn upwards instead of downwards with respect to increasing q. Naturally the anomalous dispersion could become normal by including the interplanar coupling. On the other hand, the measured density of states (Fig. 4) seems to be linear. This would be compatible with an anomalous dispersion. The fact that we do not see any soft modes in S(Q,E) (Fig. 10) demonstrates that the interplanar coupling is certainly not negligible. If this was the case, then soft modes along the stacking direction would be inevitable. Therefore, the contraction of the plane certainly should influence the physics along the stacking axis.

**B. Grüneisen parameters and thermal expansion**

The calculation of thermal expansion is carried out using the quasi-harmonic approximation. Each mode of energy E$_i$ contributes to the volume thermal expansion coefficient [31] given by $\alpha_V = \frac{1}{BV}\sum_i \Gamma_i C_{Vi}(T)$, where $V$ is the unit cell volume, $B$ is the bulk modulus, $\Gamma_i$ ( $= -\partial \ln E_i / \partial \ln V$) are the mode Grüneisen parameters and $C_{Vi}$ the specific-heat contributions of the phonons in state i (= **q**j) of energy $E_i$. Our published high-pressure inelastic neutron scattering experiments [17] on polycrystalline samples of Zn(CN)$_2$ enabled us to estimate the energy dependence of the ratios $\frac{\Gamma_i}{B}$ at 165 K and 225 K. The measurements show that the $\frac{\Gamma_i}{B}$ values are nearly the same at 165 K and 225 K. The thermal expansion coefficient derived from the phonon data is in good agreement with that obtained from diffraction measurements.

In order to estimate theoretically the volume thermal expansion coefficient, we have calculated the Grüneisen parameters based on the phonon density of states at two different unit-cell volumes. In addition to the phonon spectra at the experimental cell volumes, we calculated the phonon spectra using the lattice parameters reduced by 0.2% together with re-optimized atomic positions. The calculation of the bulk modulus, B, was then accomplished by determining the total energy of the materials as a function of unit-cell volumes and fitting them to a Birch equation of state [32]. We obtain values for B of 84.1 GPa for Zn(CN)$_2$ and 63.4 GPa for Ni(CN)$_2$.

The calculated $\frac{\Gamma_i}{B}$ for Zn(CN)$_2$ and Ni(CN)$_2$ are shown in Figs. 11 and 12 (a), respectively. The modes up to 15 meV show negative $\frac{\Gamma_i}{B}$, with the low-frequency modes around 2 meV for Zn(CN)$_2$ showing the largest negative $\frac{\Gamma_i}{B}$. The calculated $\frac{\Gamma_i}{B}$ for Zn(CN)$_2$ (Fig. 11) are in very good agreement with the values obtained from our high-pressure inelastic neutron scattering measurements and for higher frequency modes with the *ab-initio* calculations done by Zwanziger [33]. The calculated temperature dependence of the volume thermal expansion coefficient derived from these $\frac{\Gamma_i}{B}$ values compares very well with those derived from our phonon data (Fig. 13(a)). These values for Zn(CN)$_2$ can be used to calculate the volume expansion, both are in good agreement with the corresponding value obtained from diffraction data [5] (Fig. 13(b)). The *ab-initio* calculations by Zwanziger [33] give a thermal expansion coefficient of $-12 \times 10^{-6}$ K$^{-1}$ at 5 K, in agreement with the experimental data. However, Zwanziger [33] did not report Grüneisen parameters of modes below 3 meV, or a detailed temperature dependence of $\alpha_V$. Note that the value of $\alpha_V$ changes to $-51 \times 10^{-6}$ at 300 K.

Ni(CN)$_2$ shows two-dimensional NTE in the *a-b* plane with $\alpha_a$ = -6.5 × 10$^{-6}$ K$^{-1}$. The large positive expansion along *c* ($\alpha_c$= 61.8 × 10$^{-6}$ K$^{-1}$) results in an increase in volume with temperature ($\alpha_V$ =48.5 × 10$^{-6}$ K$^{-1}$). The experimental $\frac{\Gamma_i}{B}$ values are not available, however, these values should be positive for positive $\alpha_V$. We have carried out *ab-initio* calculations for Ni(CN)$_2$ on what are effectively isolated sheets. By separating the sheets, we are able to employ *P4* symmetry and achieve a great saving in computational resource. Structures with the sheets at the correct separation and alignment can only be described in *P1*. This is the case because in contrast to the modeling of diffraction patterns, where an average C/N atom can be employed, in *ab-initio* calculations discrete C or N atoms must occupy any individual atomic site. The use of *P1* symmetry would be prohibitively expensive in computing time in *ab-initio* calculations of the type employed here. Attempts to introduce stacking disorder of the type found in Ni(CN)$_2$ into the modelling would present additional computing costs.

We find that such a model gives negative $\frac{\Gamma_i}{B}$ (Fig. 12(a)) for Ni(CN)$_2$. The calculated average $\alpha_V$ (Fig. 12(b)) in the 100-300 K range is -16.5 × 10$^{-6}$ K$^{-1}$, which, gives a linear $\alpha_L$ = -5.5 × 10$^{-6}$ K$^{-1}$ and compares excellently with the $\alpha_a$ value of -6.5 × 10$^{-6}$ K$^{-1}$ from diffraction experiments [5]. Unfortunately, our modeling produces no quantitative information on the third dimension because we have modeled effectively isolated sheets. A qualitative explanation of the overall positive expansion of this system is that as these layers contract in the *a-b* plane they expand into the third dimension pushing the layers apart as suggested in Hibble *et al.* [3]. The weak interactions between the layers mean that expansion in this direction is easy and explains the overall PTE in this system.

The estimated $\frac{\Gamma_i}{B}$ values from *ab-initio* calculations (Fig. 11) have been used to estimate the contribution of the various phonons to the thermal expansion (Fig. 14) as a function of phonon energy in Zn(CN)$_2$ at 165 K. Previously, we estimated this contribution from our experimental phonon data [17]. We find that the estimation from experiment and *ab-initio* calculation gives similar results. The maximum negative contribution to $\alpha_V$ stems from the low-energy transverse acoustic modes around 2.5 meV. The modes in the vicinity of 7.5 meV also contribute substantially to NTE. Similarly for Ni(CN)$_2$, we find (Fig. 14) that maximum contribution to NTE is from phonon modes of around energy 2.5 meV.

In order to determine the character of phonon modes in the two compounds, we have calculated the mean squared displacements of various atoms, $<(u^2)>$, arising from all phonons of energy E, as follows

$$<(u^2)_k> = (n+\frac{1}{2})\frac{\hbar}{m_k \omega_k} \quad (3)$$

where $n = \left[\exp(\frac{\hbar \omega_k}{k_B T}) - 1\right]^{-1}$, $\omega_k$ and $m_k$ are the mode frequency and mass of the $k^{th}$ atom in the unit cell respectively.

The calculated partial density of states (Fig. 6) has been used for this calculation. Equal contributions for all the atoms (Fig. 15) up to 5 meV in Zn(CN)$_2$ and 12 meV in Ni(CN)$_2$ indicate predominantly acoustic modes. The contributions from Zn and Ni atoms vanish above these energies. It is interesting to note that the amplitudes are weighted with the inverse masses while this is not the case for the partial densities of states, where Zn and Ni contributions are visible up to rather high frequencies (see Fig. 6). It is reasonable to conclude that the lowest bands without significant Zn contribution to the vibration amplitude correspond to rotational modes in Zn(CN)$_2$. The ZnC$_4$ and ZnN$_4$ rigid-unit modes in Zn(CN)$_2$ are therefore found at lower energies (5 meV to 12 meV), in comparison to NiC$_4$ and NiN$_4$ (12 meV to 16 meV) rigid-unit modes in Ni(CN)$_2$.

## C. Anharmonicity

The analysis presented here is important for understanding the anharmonic nature of phonons in these compounds. The change in phonon energies with temperature is due to implicit as well as explicit anharmonicities. The implicit anharmonicity of phonons is due to the change of the unit-cell volume and/or concomitant changes of structural parameters with temperature. The explicit anharmonicity includes changes in phonon frequencies due to large thermal amplitude of atoms. The temperature-dependent measurements (total anharmonicity) include both effects. The present measurements of phonon spectra along with the previously reported pressure dependence of the phonon spectra (implicit anharmonicity) can be used to separate [34] the temperature effect at constant volume (explicit or true anharmonicity) as

$$\left.\frac{dE_i}{dT}\right|_P = \left.\frac{\partial E_i}{\partial T}\right|_V + \left.\frac{\partial E_i}{\partial V}\right|_T \cdot \left.\frac{\partial V}{\partial T}\right|_P$$

Using $\Gamma_i = -\frac{V}{E_i}\left.\frac{\partial E_i}{\partial V}\right|_T$ and $\alpha = -\frac{1}{V}\left.\frac{\partial V}{\partial T}\right|_P$, one obtains

$$\left.\frac{1}{E_i}\frac{dE_i}{dT}\right|_P = \left.\frac{1}{E_i}\frac{\partial E_i}{\partial T}\right|_V - \Gamma_i\cdot\alpha \qquad (4)$$

Here the first term on the right hand side is the true anharmonic (explicit) contribution, and the second the quasiharmonic (implicit) term. The left side term represent the total anharmonicity. The temperature dependence of phonon spectra has been used for estimating the total anharmonicity $\left(\left.\frac{1}{E_i}\frac{dE_i}{dT}\right|_P\right)$ of both compounds. For Zn(CN)$_2$, the $\left.\frac{1}{E_i}\frac{dE_i}{dT}\right|_P$ values for phonons of energy $E_i$ have been obtained using the cumulative distributions of the experimental data of phonon density of states at 180 K and 240 K, while for Ni(CN)$_2$, the experimental data at 160 K and 220 K has been used to obtain $\left.\frac{1}{E_i}\frac{dE_i}{dT}\right|_P$.

As mentioned above, the temperature-dependent measurements give estimates of the total anharmonicity of phonons and include both the implicit and explicit effects. On increasing the temperature, the implicit anharmonicity results in a decrease of phonon frequencies for all materials irrespective of their thermal expansion coefficients. However, explicit contribution may cause either an increase or decrease of phonon frequencies with increasing temperature. Finally, it is the net sum of the two components, which we have observed in the measurements and shown in Figs. 3 and 4. We find that for Ni(CN)$_2$ the phonon energies increases with increase (Fig. 16) of temperature, hence the total anharmonicity $\left(\left.\frac{1}{E_i}\frac{dE_i}{dT}\right|_P\right)$ is positive. In particular, modes below 2 meV have very large total anharmonicity. Since the implicit part would result in softening of modes with increase of temperature, the hardening of modes with increase of temperature gives us evidence for the large explicit anharmonic nature of phonons in Ni(CN)$_2$.

In the case of Zn(CN)$_2$ (Fig. 16), for modes below 2 meV, the total anharmonicity of phonons is negative, while it is positive for high-energy modes. Since the implicit part causes decrease of phonon frequencies with increase of temperature, the hardening of modes above 2 meV must be due to large positive explicit anharmonicity of these phonons.

The present temperature dependence of phonon spectra as well as our previous measured pressure dependence can be used for separating the quasiharmonic and true anharmonicity in $Zn(CN)_2$. The total anharmonicity of phonon modes in $Zn(CN)_2$ is shown in Fig. 17. For materials with negative or positive thermal expansion coefficients ($\alpha$), the second term on the right-hand side $(\Gamma_i.\alpha)$ is always positive. $Zn(CN)_2$ has negative thermal expansion over its entire temperature range of stability. The quasiharmonic contribution obtained from our previous high-pressure density of states measurements is shown in Fig. 17. The explicit $\left(\frac{1}{E_i}\frac{dE_i}{dT}\bigg|_V\right)$ contribution for phonon modes up to 30 meV in $Zn(CN)_2$ extracted using equation (4) is shown in Fig. 17. The magnitude of anharmonicities of modes in these compounds is substantially larger than those in other typical solids [35] where it ranges between $2 \times 10^{-5}$ K$^{-1}$ and $10 \times 10^{-5}$ K$^{-1}$.

## IV. CONCLUSIONS

We have reported measurements of the temperature dependence of phonon spectra for $Zn(CN)_2$ and $Ni(CN)_2$ and results of *ab initio* lattice dynamical calculations. The comparison of phonon spectra for Zn and Ni compounds show strong renormalization effects in the phonon spectra of these compounds, which cannot be simply explained by the lattice contraction and mass effect. The phonon spectra have been well reproduced by using *ab-initio* calculations. The anomalous thermal expansion bahaviour in both the compounds have been estimated. Our calculated NTE coefficient in $Zn(CN)_2$ agrees nicely with the experimental data. Calculations show that phonon modes of energy about 2 meV are major contributors to NTE. The measured temperature dependence of the phonon spectra along with our previous pressure-dependent phonon measurements has been used for estimating the quasiharmonic and true anharmonicity. The value for the NTE coefficient in the plane of the layered material $Ni(CN)_2$ has been calculated and found to be in excellent agreement with that determined experimentally. We have shown that low energy phonon modes in these compounds are strongly anharmonic.

## ACKNOWLEDGEMENT

The University of Reading is thanked for the provision of the Chemical Analysis Facility.

[1] T. A. Mary, J. S. O. Evans, T. Vogt and A. W. Sleight, Science **272**, 90 (1996).

[2] C. Lind, A. P. Wilkinson, Z. Hu, S. Short, and J. D. Jorgensen, Chem. Mater. **10**, 2335 (1998).

[3] S. J. Hibble, A. M. Chippindale, A. H. Pohl, A. C. Hannon, Angew. Chem. Int. Ed. **46**, 7116 (2007).

[4] A. L. Goodwin, M. T. Dove, A. M. Chippindale, S. J. Hibble, A. H. Pohl and A. C. Hannon, Phys. Rev. **B80**, 054101 (2009).

[5] A. L. Goodwin, C. J. Kepert, Phys. Rev. **B71**, R140301 (2005).

[6] A. L. Goodwin, M. Calleja, M. J. Conterio, M. T. Dove, J. S. O. Evans, D. A. Keen, L. Peters, and M. G. Tucker, Science **319**, 794 (2008).

[7] K. W. Chapman, P. J. Chupas, and C. J. Kepert, J. Am. Chem. Soc. **127**, 15630 (2005).

[8] P. Ding, E. J. Liang, Y. Jia and Z. Y. Du, J. Phys.: Condens. Matter **20**, 275224 (2008).

[9] D. J. Williams, D. E. Partin, F. J. Lincoln, J. Kouvetakis, M. O'Keefe, J. Solid State Chem. **134**, 164 (1997).

[10] W. I. David, J. S. O. Evans and A. W. Sleight, Europhysics Letters **46**, 661 (1999).

[11] G. Ernst, C. Broholm, G. R. Kowach and A. P. Ramirez, Nature **396**, 147 (1998).

[12] Y. Yamamura, N. Nakajima, T. Tsuji, M. Koyano, Y. Iwasa, S. Katayama, K. Saito, and M. Sorai, Phys. Rev. **B66**, 014301 (2002).

[13] R. Mittal, S. L. Chaplot, H. Schober and T. A. Mary, Phys. Rev. Lett. **86**, 4692 (2001).

[14] R. Mittal, S. L. Chaplot, A. I. Kolesnikov, C. -K. Loong and T. A. Mary, Phys. Rev. **B68**, 54302 (2003).

[15] R. Mittal and S. L. Chaplot, Phys. Rev. **B60**, 7234 (1999).

[16] K. W. Chapman, M. Hagen, C. J. Kepert and P. Manuel, Physica **B 385-386**, 60 (2006).

[17] R. Mittal, S. L. Chaplot and H. Schober, Appl. Phys. Lett. **95**, 201901 (2009).

[18] D. L. Price and K. Skold, in *Neutron Scattering*, edited by K. Skold and D. L. Price (Academic Press, Orlando, 1986), Vol. A; J. M. Carpenter and D. L. Price, Phys. Rev. Lett. **54**, 441 (1985).

[19] www.ncnr.nist.gov; V. F. Sears, Neutron News **3**, 29 (1992); A. -J. Dianoux and G. Lander (Eds.), *Neutron Data Booklet, Institut Laue-Langevin*, Grenoble, France (2002).

[20] P. E. Blöchl, Phys. Rev. **B50**, 17953 (1994).

[21] P. Hohenberg and W. Kohn, Phys. Rev. **136**, 864 (1964).

[22] W. Kohn and L. J. Sham, Phys. Rev. **140**, 1133 (1965).

[23] G. Kresse and J. Furthmüller, Comput. Mater. Sci. **6**, 15 (1996).

[24] G. Kresse and D. Joubert, Phys. Rev. **B59**, 1758 (1999).

[25] J. P. Perdew, K. Burke and M. Ernzerhof, Phys. Rev. Lett. **77**, 3865 (1996).

[26] J. P. Perdew, K. Burke and M. Ernzerhof, Phys. Rev. Lett. **78**, 1396 (1997).

[27] K. Parlinski, Z.-Q. Li, and Y. Kawazoe, Phys. Rev. Lett. **78**, 4063 (1997).

[28] K. Parlinksi, Software phonon, 2003.

[29] R. Mittal, S. Mitra, H. Schober, S. L. Chaplot, R. Mukhopadhyay, arXiv:0904.0963.

[30] T. R. Ravindran, A. K. Arora, Sharat Chandra, M. C. Valsakumar and N. V. Chandra Shekar, Phys. Rev. **B76**, 54302 (2007).

[31] G. Venkataraman, L. Feldkamp and V.C. Sahni, *Dynamics of perfect crystals*, MIT Press, Cambridge (1975).

[32] F. Birch, J. Geophys. Res. **57**, 227 (1952).

[33] J. W. Zwanziger, Phys. Rev. **B76**, 052102 (2007).

[34] B. A. Weinstein and R. Zallen, in *Light Scattering in Solids IV*, edited by M. Cardona and G. Guntherodt (Springer-Verlag, Berlin, 1984), p 463.

[35] R. Zallen and E. M. Conwell, Solid State Commun. **31**, 557 (1979).

[36] K. W. Chapman and P. J. Chupas, J. Am. Chem. Soc. **129**, 10090 (2007).

TABLE I. Fractional atomic coordinates used to generate the Ni(CN)$_2$ layers within the tetragonal space group *P4* for Ni(CN)$_2$. $a = b = 6.86900$ Å and $c = 6.40500$ Å

| | | | |
|---|---|---|---|
| Ni | 0.0 | 0.0 | 0.0 |
| Ni | 0.5 | 0.5 | 0.0 |
| C | 0.1909 | 0.1909 | 0. |
| N | 0.3091 | 0.3091 | 0.0 |

TABLE II. *Ab-initio* calculated (Calc) and observed (Exp) [30] Raman and IR frequencies (cm$^{-1}$) for Zn(CN)$_2$. Irrep, Type and M stand for irreducible representation, type of the mode and multiplicity, respectively. R, RI and S indicate if the mode is Raman active or is both Raman and IR active or optically inactive, respectively. The point group symmetry is $T_d^1$.

| Calc | 59 | 173 | 178 | 204 | 240 | 330 | 336 | 476 | 481 | 2251 | 2261 |
|---|---|---|---|---|---|---|---|---|---|---|---|
| Exp | | | 178 | 216 | | 343 | 339 | 461 | | 2218 | 2221 |
| Irrep | T$_1$ | E | T$_2$ | T$_2$ | T1 | T$_2$ | E | A$_1$ | T$_2$ | T$_2$ | A$_1$ |
| Type | S | R | RI | RI | S | RI | R | RI | R | RI | R |
| M | 3 | 2 | 3 | 3 | 3 | 3 | 2 | 3 | 1 | 3 | 1 |

TABLE III. *Ab-initio* calculated (Calc) and observed (Exp) Raman and IR frequencies (cm$^{-1}$) for Ni(CN)$_2$. Irrep, Type and M stand for irreducible representation, type of the mode and multiplicity, respectively. R and RI indicate if the mode is Raman active or is both Raman and IR active, respectively. The point group symmetry is $C_4^1$, thus all IR are Raman active. A and E Irreps (polar modes) are also IR, with polarizations lying along the *z*-axis and in the *xy*-plane, respectively. The B modes are Raman active only.

| Calc | 99 | 100 | 103 | 210 | 303 | 328 | 333 | 334 | 335 | 337 |
|---|---|---|---|---|---|---|---|---|---|---|
| Exp | | | | 200 | | | | 334 (broad) | | |
| Irrep | E | A | A | B | E | E | A | A | B | E |
| Type | RI | RI | RI | R | RI | RI | RI | RI | R | RI |
| M | 2 | 1 | 1 | 1 | 2 | 2 | 1 | 1 | 1 | 2 |
| Calc | 397 | 461 | 489 | 490 | 566 | 583 | 606 | 2196 | 2205 | 2238 |
| Exp | | | | 508 | 561 | 604 | 607 | 2202 | 2206 | 2215 |
| Irrep | B | A | E | B | B | A | E | E | B | A |
| Type | R | RI | RI | R | R | RI | RI | RI | R | RI |
| M | 1 | 1 | 2 | 1 | 1 | 1 | 2 | 2 | 1 | 1 |

FIG. 1. The structure of Zn(CN)$_2$ in *P43m*. Key: Zn, grey spheres; C, green spheres; N, blue spheres

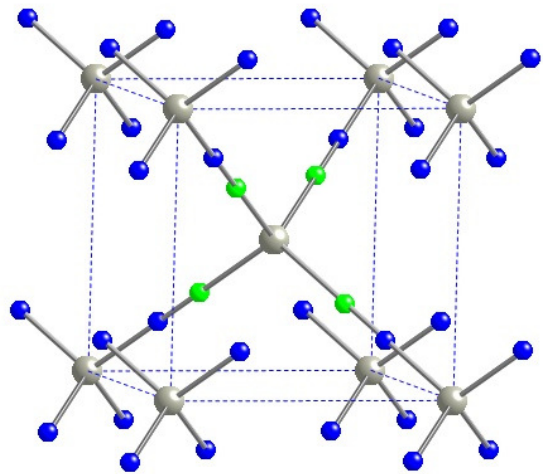

FIG. 2 The structure of one layer of Ni(CN)$_2$ with $D_{4h}$ symmetry. Key: Ni, grey spheres; C, green spheres; N, blue spheres

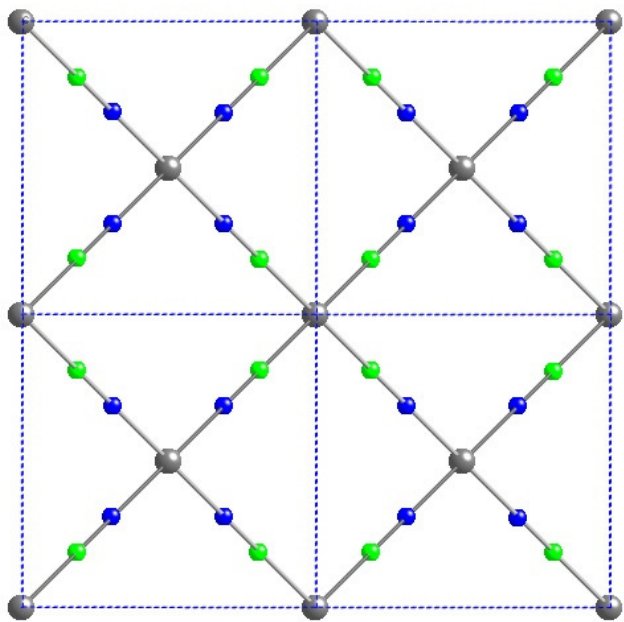

FIG. 3. (Color online) The temperature dependence of the phonon spectra for $Zn(CN)_2$. The phonon spectra are measured with incident neutron wavelength of 5.12 Å using the IN6 spectrometer at ILL. The calculated phonon spectra from *ab-initio* calculations are also shown. The calculated spectra have been convoluted with a Gaussian of FWHM of 10% of the energy transfer in order to describe the effect of energy resolution in the experiment carried out using the IN6 spectrometer.

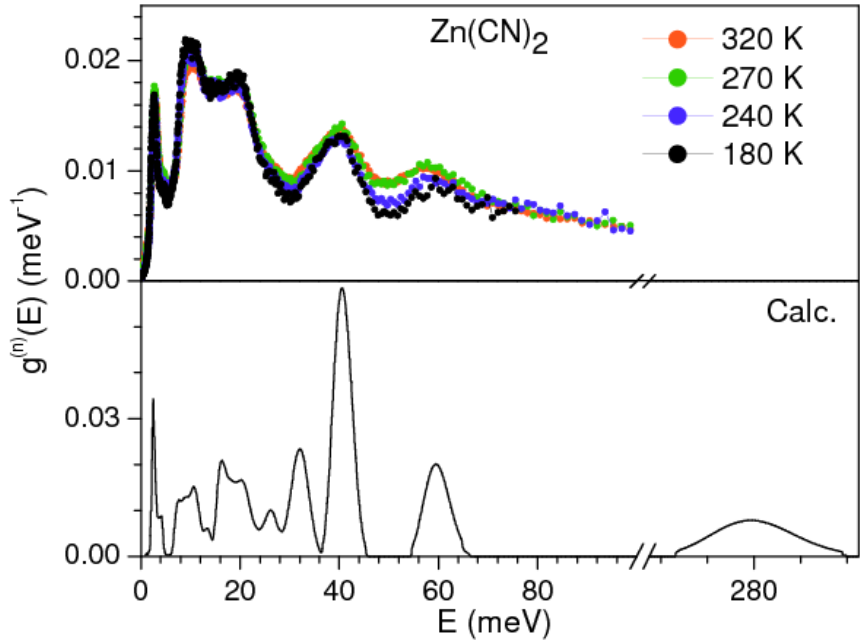

FIG. 4. (Color online) The temperature dependence of the phonon spectra for $Ni(CN)_2$. The phonon spectra are measured with incident neutron wavelength of 4.14 Å using the IN6 spectrometer at ILL. The calculated phonon spectra from *ab-initio* calculations are also shown. The calculated spectra have been convoluted with a Gaussian of FWHM of 10% of the energy transfer in order to describe the effect of energy resolution in the experiment carried out using the IN6 spectrometer.

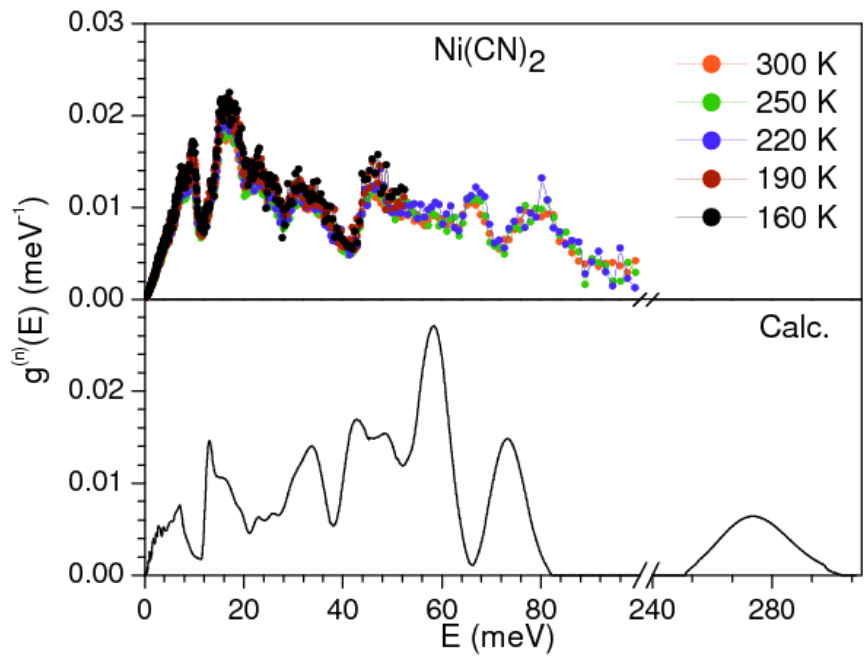

FIG. 5. (Color online) Comparison of the experimental phonon spectra for Zn(CN)$_2$ and Ni(CN)$_2$.

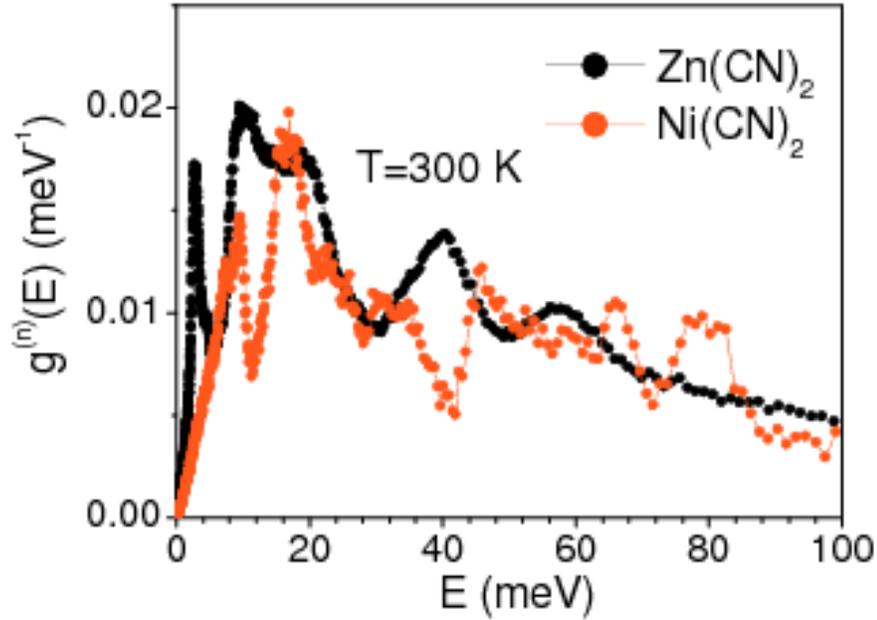

FIG. 6. (Color online) The calculated partial density of states for the various atoms in Zn(CN)$_2$ and Ni(CN)$_2$.

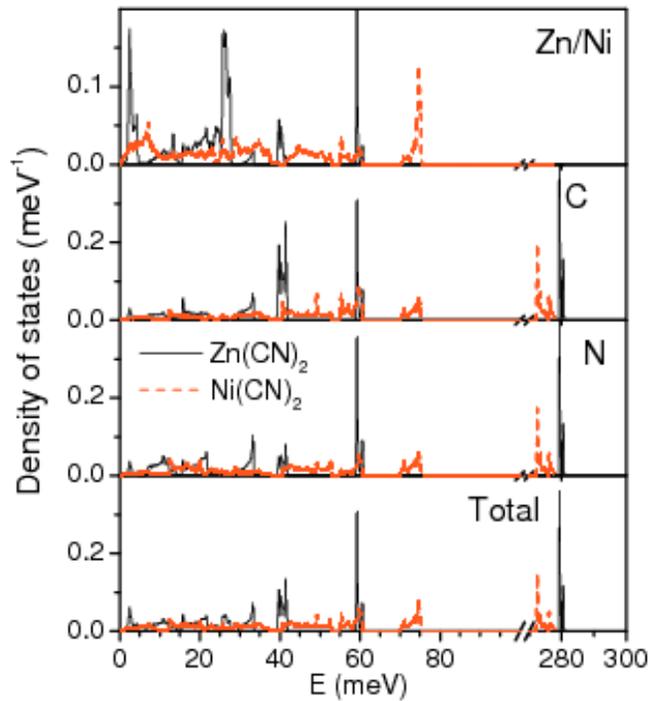

FIG. 7 Raman spectrum of Ni(CN)$_2$. Inset shows enlargement of the low-wavenumber region.

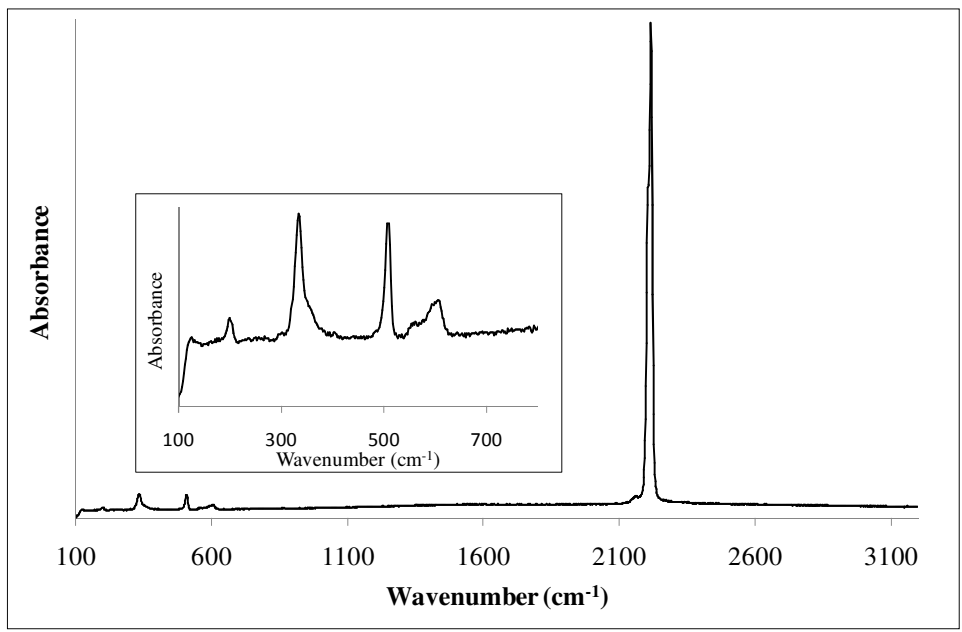

FIG. 8. The $\nu_{C\equiv N}$ region of the Raman (black) and infrared (grey) spectra for Ni(CN)$_2$.

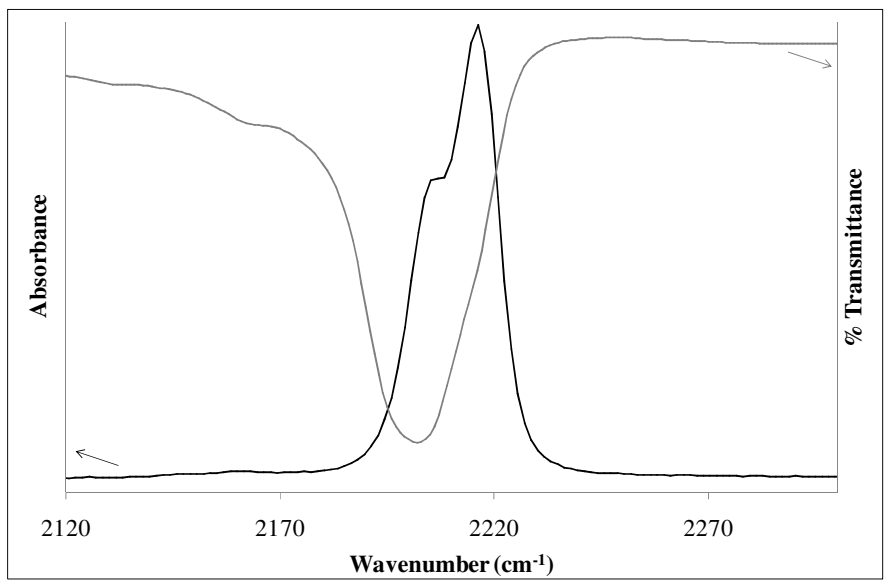

FIG. 9. The calculated phonon dispersion curves for Zn(CN)$_2$ and Ni(CN)$_2$. The Bradley-Cracknell notation is used for the high-symmetry points along which the dispersion relations are obtained. Zn(CN)$_2$: Γ=(0,0,0); X=(1/2,0,0); M=(1/2,1/2,0) and R=(1/2,1/2,1/2). Ni(CN)$_2$: Γ=(0,0,0); X(1/2,0,0) and M(1/2,1/2,0). In order to expand the *y*-scale, the set of four and three number of dispersionless modes respectively in Zn(CN)$_2$ and Ni(CN)$_2$ due to the cyanide stretch at about 280 meV are not shown.

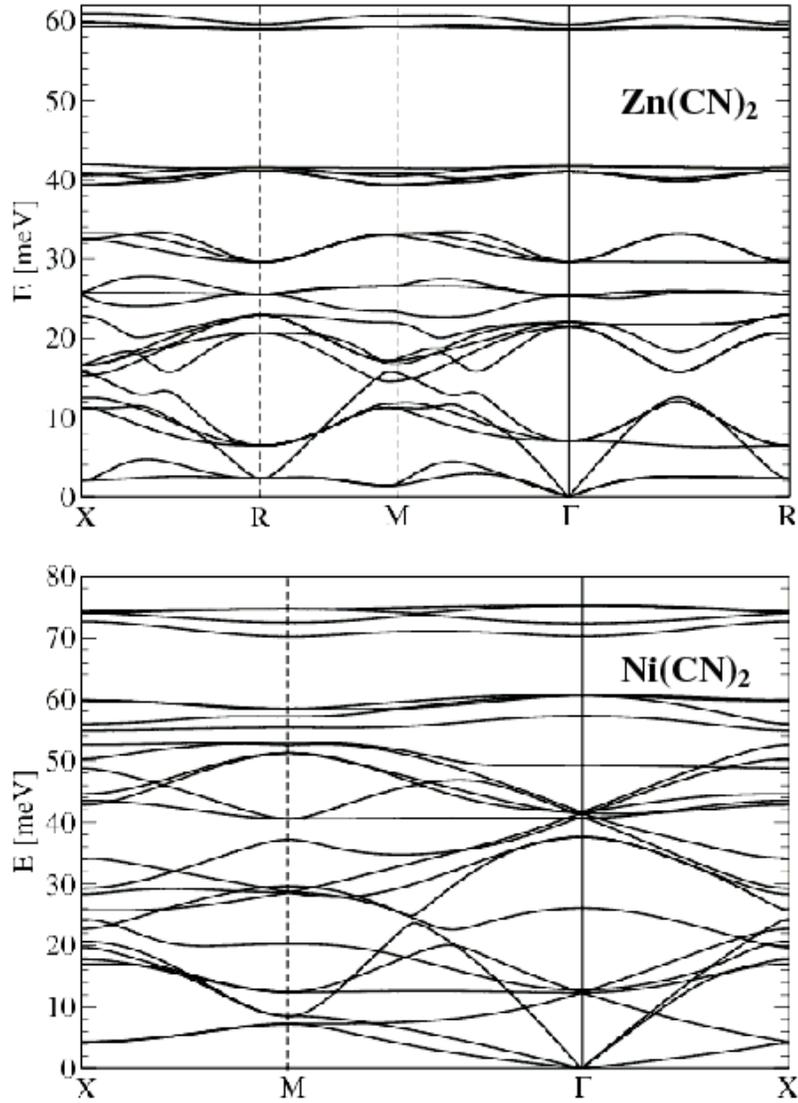

FIG. 10. (Color online) The experimental Bose-factor corrected S(Q,E) plots for Zn(CN)$_2$ and Ni(CN)$_2$ at 180 K and 160 K respectively. The values of S(Q,E) are normalized to the mass of sample in the beam. For clarity, a logarithmic representation is used for the intensities. The measurements for Zn(CN)$_2$ and Ni(CN)$_2$ were performed with an incident neutron wavelength of 5.12 Å (3.12 meV) and 4.14 Å (4.77 meV) respectively

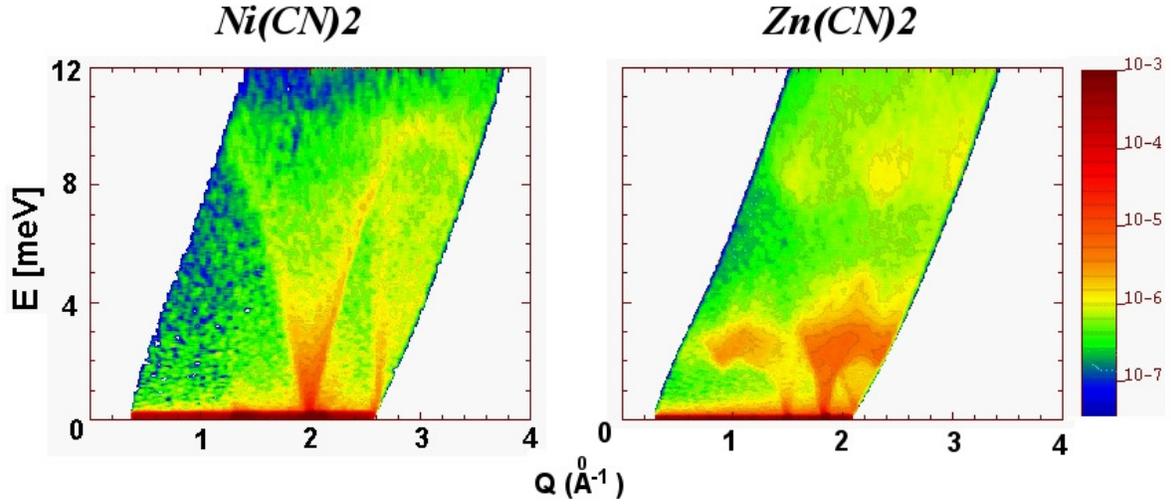

FIG. 11. (Color online) The comparison between the experimental and calculated $\frac{\Gamma_i}{B}$ as a function of phonon energy $E$. The $\frac{\Gamma_i}{B}$ values derived from *ab-initio* calculations from Ref. [33] are shown by closed circles. The $\frac{\Gamma_i}{B}$ values represent the average over the whole Brillouin zone.

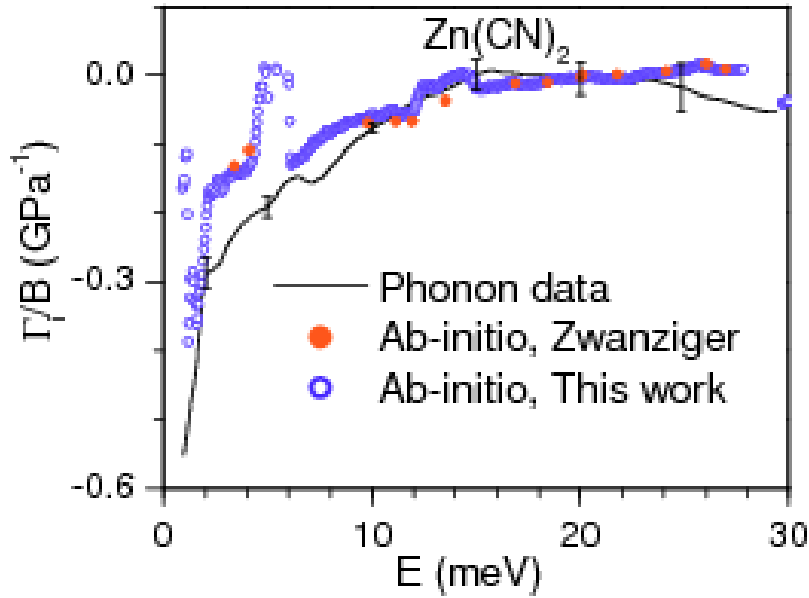

FIG. 12. (a) The calculated $\frac{\Gamma_i}{B}$ and (b) volume thermal expansion coefficient ($\alpha_V$) derived for Ni(CN)$_2$ from *ab-initio* calculations.

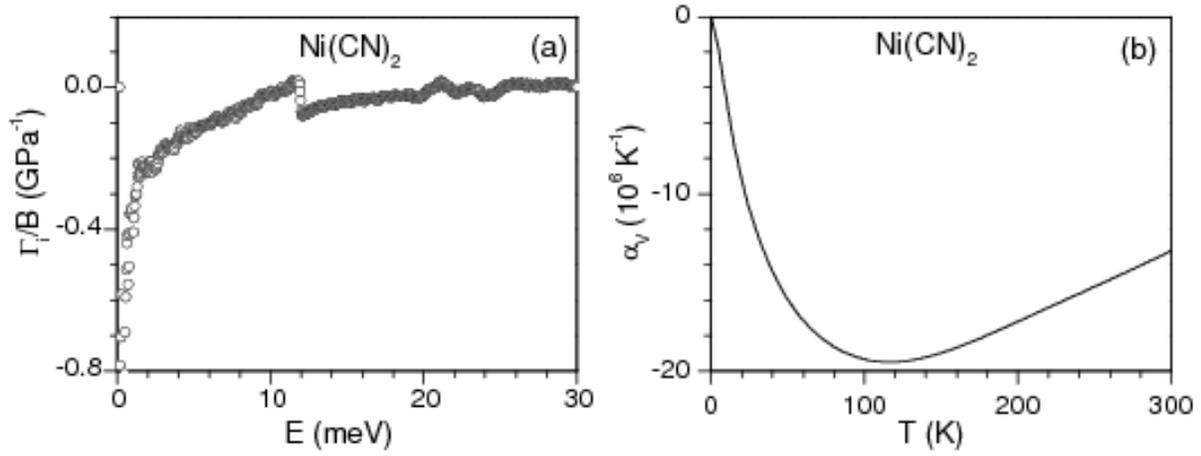

FIG. 13. (Color online) (a) The comparison between the volume thermal expansion coefficient ($\alpha_V$) derived from the *ab-initio* calculations and experimental $\frac{\Gamma_i}{B}$ values [17] at 165 K. (b) The comparison between the volume thermal expansion derived from the present *ab-initio* calculations (solid line), high-pressure inelastic neutron scattering experiment (dashed line) and that obtained using X-ray diffraction [5] (open circles).

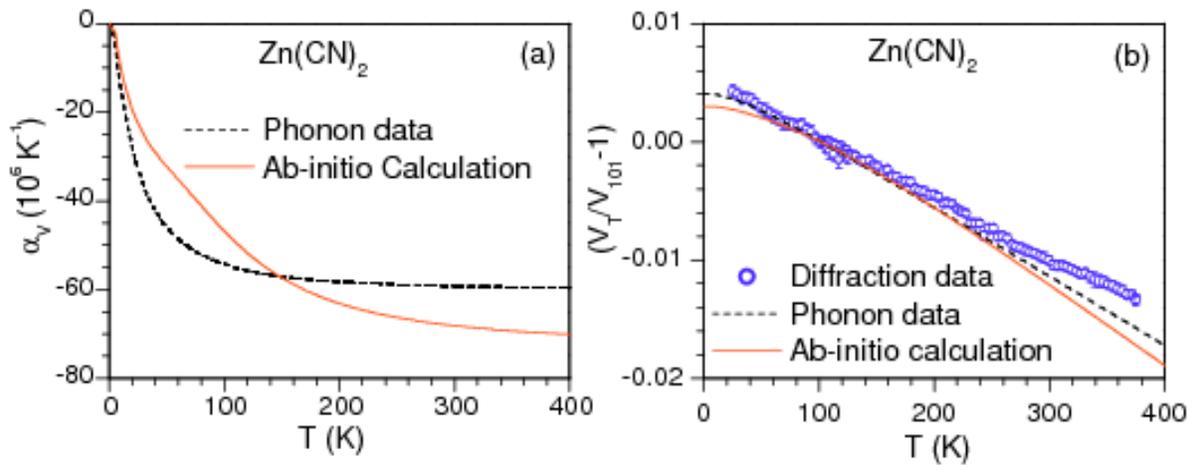

FIG. 14 The contribution of phonons of energy $E$ to the volume thermal expansion coefficient ($\alpha_V$) as a function of $E$ at 165 K in $Zn(CN)_2$ and $Ni(CN)_2$. The experimental phonon data for $Zn(CN)_2$ are taken from Ref. [17].

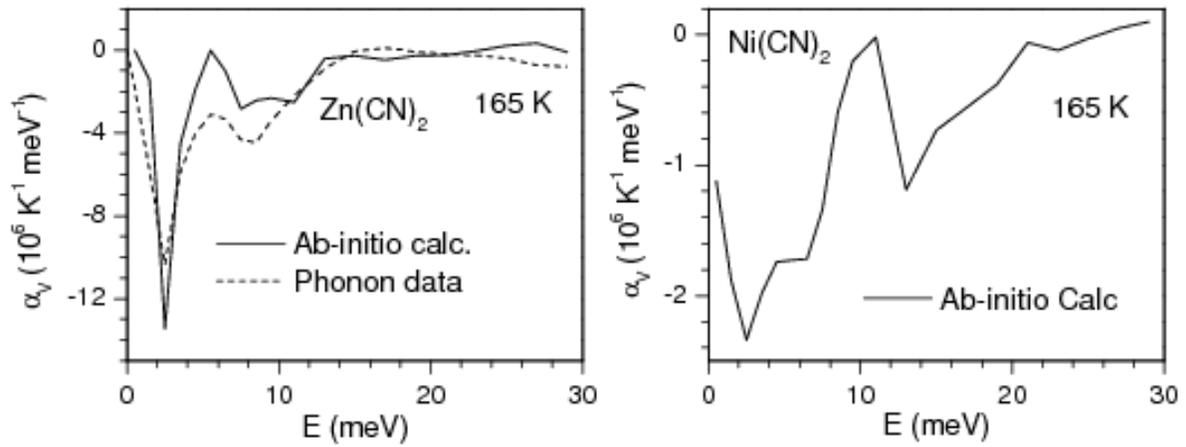

FIG. 15. (Color online) The calculated contribution to the mean squared amplitude of the various atoms arising from phonons of energy E at $T=300$ K in $Zn(CN)_2$ and $Ni(CN)_2$.

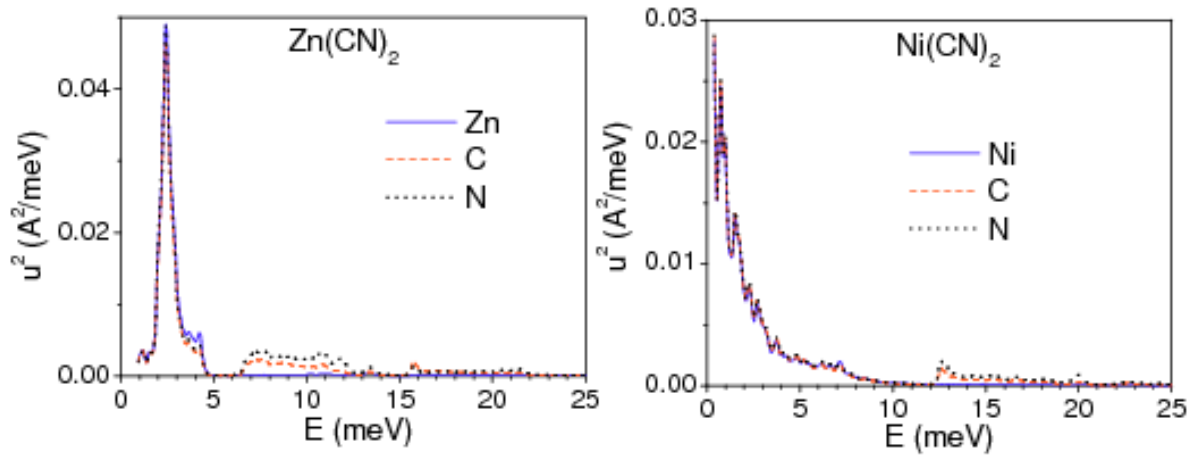

FIG. 16. (Color online) The total anharmonicities $\left(\frac{1}{E_i}\frac{dE_i}{dT}\bigg|_P\right)$ of different phonons of energy E in Zn(CN)$_2$ and Ni(CN)$_2$. The $\frac{1}{E_i}\frac{dE_i}{dT}\bigg|_P$ has been obtained using the cumulative distributions of the experimental data of phonon density of states of Zn(CN)$_2$ at 180 K and 240 K, while for Ni(CN)$_2$ the experimental data at 160 K and 220 K have been used.

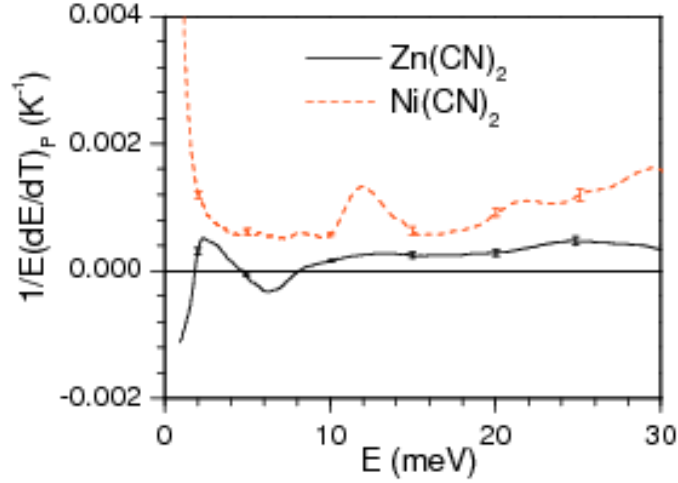

FIG. 17. (Color online) Total anharmonicity, implicit (quasiharmonic) and explicit (true anharmonic) contributions of different phonons in Zn(CN)$_2$. The total anharmonicity $\left(\frac{1}{E_i}\frac{dE_i}{dT}\bigg|_P\right)$ has been obtained using the cumulative distributions of the experimental data of phonon density of states of Zn(CN)$_2$ at 180 K and 240 K. The quasiharmonic contribution has been obtained [17] using the pressure dependence of phonon density of states at 165 K. The bulk modulus B value of 34.19 GPa [36] has been used for estimating $\Gamma_i$ values from our measured [17] $\frac{\Gamma_i}{B}$ values.

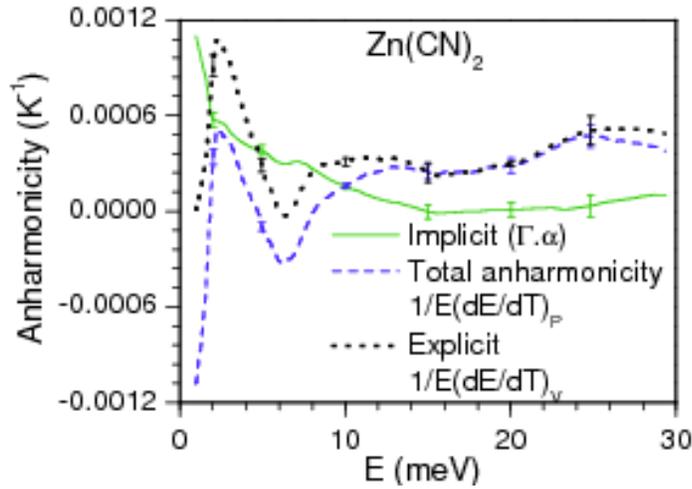